\documentclass[fleqn,twoside]{article}
\usepackage{espcrc2}

\usepackage{graphicx}
\usepackage[figuresright]{rotating}

\newcommand{\AmS}{{\protect\the\textfont2
  A\kern-.1667em\lower.5ex\hbox{M}\kern-.125emS}}

\title{
\thispagestyle{empty}
\vspace{-25mm}
\rightline{\small DESY 03-144~~~~~}
\rightline{\small ITEP-LAT/2003-25~~~~~}
\rightline{\small KANAZAWA-03-29~~~~~}
\vspace{10mm}
Structure of the baryonic flux tube in $N_{f}=2$ lattice QCD at finite temperature.
\thanks{Talk given by Y.~Mori. at Lattice'03.}
}
\author{
Y. Mori\address[KITP]{Institute for Theoretical Physics, Kanazawa University, Kanazawa 920-1192, Japan\\[-0.5em]}, 
V. Bornyakov\address{Institute for High Energy Physics, RU-142284 Protvino, Russia\\[-0.5em]}, 
H. Ichie\addressmark,
Y. Koma\address[MPI]{Max-Planck-Institut f\"ur Physik, D-80805 M\"unchen, Germany\\[-0.5em]},
M. Polikarpov\address{ITEP, B.Cheremushkinskaya 25, RU-117259 Moscow, Russia\\[-0.5em]}, 
G. Schierholz\address[DASY]{NIC/DESY Zeuthen, Platanenallee 6, 15738 Zeuthen, Germany and Deutsches Elektronen-Synchrotron DESY D-22603 Hamburg, Germany\\[-0.5em]}, 
 H.~St\"uben\address{Konrad-Zuse-Zentrum f\"ur Informationstechnik
Berlin, D-14195 Berlin, Germany\\[-0.5em]} and
T.~Suzuki$^{\rm a}$
}

\begin{document}
\begin{abstract}
We study the flux tube profile in the baryonic system in full QCD at finite temperature on $N_{t}=8$ lattice.
We fix the maximally Abelian gauge
and measure the monopole and the photon parts of the Abelian action density,
the color electric field and the monopole current on both sides of the finite temperature transition.
We demonstrate the disappearance of the flux tube in the high temperature phase.
\vspace{-1pc}
\end{abstract}

\maketitle

\section{INTRODUCTION}
Lattice studies of the baryonic system consisting of three static quarks (3Q) are important for clarification of the baryon structure.
Recently there appeared a number of papers devoted to the studies of the 3Q system at zero temperature 
\cite{Alexandrou_2002,Takahashi_2002,Ichie:2002mi,Ichie:2002dy}.
It was concluded that the flux tube has a  Y-shape, at least at large distances.

In this paper, we study the flux tube profile in the baryonic system at finite temperature in QCD with dynamical fermions.
The structure of the flux tube is investigated with the help of Abelian observables after the maximally Abelian gauge is fixed.
In particular, we study the monopole and the photon parts of Abelian flux tube profiles.
In our previous study of the baryonic flux tube at zero temperature
in both SU(3) gluodynamics and QCD with dynamical quarks 
\cite{Ichie:2002mi,Ichie:2002dy}
it has been demonstrated with an accuracy unmatched by  gauge invariant 
nonabelian observables that the flux tube is of Y-shape. 
The Abelian and the monopole dominance phenomena well established in the gluodynamics \cite{Suzuki:1989gp,Bali:1996dm} and in QCD with dynamical quarks \cite{DIK_2002} gives confidence that our results provide a qualitatively correct
description of the baryonic flux tube at finite temperature.
Our study is also important to check the dual superconductor scenario of confinement which predicts that the effective infrared theory of QCD should be a kind of the dual Abelian Higgs model.
\vspace{-0mm}
\section{SIMULATION DETAILS}
To study QCD with dynamical quarks we consider $N_{f}=2$ flavors of degenerate quarks, using the Wilson gauge field action and non-perturbatively  ${\cal O}(a)$ improved Wilson fermions~\cite{Booth:2001qp}.
Configurations are generated on the $16^3\,8$ lattice at $\beta=5.2$, $0.1330\le \kappa \le 0.1360$, corresponding to temperatures below and above the finite temperature transition at $\kappa_{T}=0.1344$ ($T_{c}=213(10)$~MeV)~\cite{DIK_2003}.
Details of the simulation can be found in~\cite{DIK_2003}.
We fixed the maximally Abelian (MA) gauge on generated configurations employing the simulated annealing algorithm \cite{Bali:1996dm}.
The Abelian projection procedure \cite{MaA} defines the diagonal link matrices
\[ u(s,\mu)=\mbox{diag} \{u_1(s,\mu), u_2(s,\mu),u_3(s,\mu)\}\,, \]
\[ u_l(s,\mu)=e^{i\theta_l(s,\mu)}.\]

We study the Abelian action density
\begin{equation}
\rho_{\rm ab}(s) = \frac{\beta}{3} \sum_{\mu>\nu}\sum_l
u_l(s,\mu\nu)\, ,
\label{action}
\end{equation}
where $u_l(s,\mu\nu)$ is the Abelian plaquette variable,
the Abelian color electric field
\begin{equation}
E_l(s,j)=i\bar{\theta}_l(s,j4),
\end{equation}
where $\bar{\theta}_l(s,\mu\nu)$ is the regular part of the Abelian plaquette angle, 
$\theta_l(s,\mu\nu) = \bar{\theta}_l(s,\mu\nu) + 2\pi n_l(s,\mu\nu)$, 
and the monopole current 
\begin{equation}
k_l(s,\mu)=-\frac{i}{4\pi}\epsilon_{\mu\nu\rho\sigma}\partial_{\nu}\overline{\theta}_l(s+\hat{\mu},\rho\sigma).
\end{equation}
Correlators with the baryonic source  are defined as follows:
\begin{eqnarray}
\label{oper-def}
\langle \rho_{\rm ab}(s) \rangle_{3Q} = \frac{\langle \rho_{\rm ab}(s) {\cal P}_{3Q} \rangle}{\langle {\cal P}_{3Q} \rangle}\, ,
\end{eqnarray}
where 
\begin{equation}
{\cal P}_{3Q}=\frac{1}{3!} |\varepsilon_{lmn}|\,P_l(\vec{s}_{1})P_m(\vec{s}_{2})P_n(\vec{s}_{3})
\end{equation}
\[ P_l(\vec{s}) = \prod_{t=1}^{N_t} u_l(\vec{s},t,4)\,, \]
\begin{equation}
\langle E(s,j) \rangle_{3q} = 
\label{elfield}
\end{equation}
\[ \frac{\langle \frac{1}{3!} |\varepsilon_{lmn}|
E_l(s,j) P_l(\vec{s}_1) P_m(\vec{s}_2) P_n(\vec{s}_3)
\rangle} {\langle {\cal P}_{3Q} \rangle}\,, \]
\begin{equation} 
\langle k(s,j) \rangle_{3q} = 
\label{mcurrent}
\end{equation}
\[ \frac{\langle \frac{1}{3!} |\varepsilon_{lmn}|
k_l(s,j) P_l(\vec{s}_1) P_m(\vec{s}_2) P_n(\vec{s}_3)
\rangle} {\langle {\cal P}_{3Q} \rangle}\,. \]

\begin{figure}[thpb]
\begin{center}
\includegraphics[width=60mm]{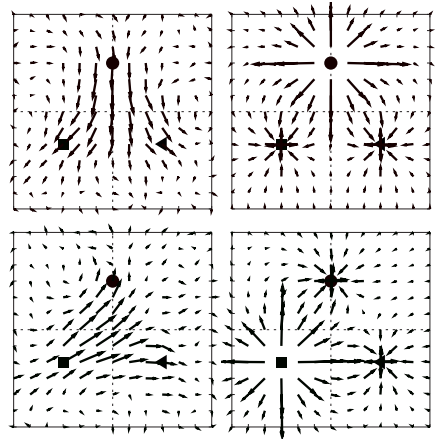}
\end{center}
\vspace{-12mm}
\caption{{\small \textit{The monopole (left column) and photon (right column) 
parts of the 
color electric field at $T/T_{c}=0.87$. The color index of the color electric field (see
eq.(\ref{elfield})) coincides with that of the topmost quark (top row) or of the leftmost
quark (bottom row).
}}}
\label{fig:profile_E.eps}
\vspace{-6mm}
\end{figure}
The Abelian field can be decomposed into the monopole and photon parts ~\cite{ploop}, and the monopole and photon observables can be defined similarly to eqs.(\ref{action})-(\ref{mcurrent}).

\section{NUMERICAL RESULTS}
First we discuss the structure of the baryonic flux tube in the confinement phase.
Fig.\ref{fig:profile_E.eps} shows the monopole and photon parts of the color electric field in the 3Q system.
The monopole part is squeezed into a flux tube while the photon part is of Coulombic form.
In the monopole component the flux tube is compatible with Y-shape.
We expect that the agreement with Y-shape will be better when the distance between quarks will be large in comparison with the intrinsic width of the flux tube.
\begin{figure}[thpb]
\begin{center}
\includegraphics[width=80mm]{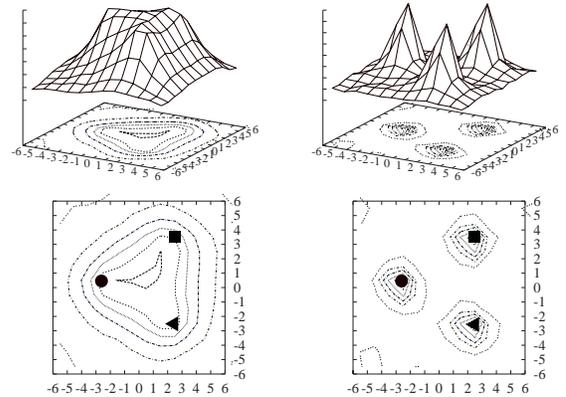}
\end{center}
\vspace{-11mm}
\caption{{\small \textit{Monopole (left) and photon (right) parts of the action 
density at $T/T_{c}=0.94$.}}}
\label{fig:profile_S.eps}
\vspace{-8mm}
\end{figure}
\begin{figure}[thpb]
\begin{center}
\includegraphics[width=70mm]{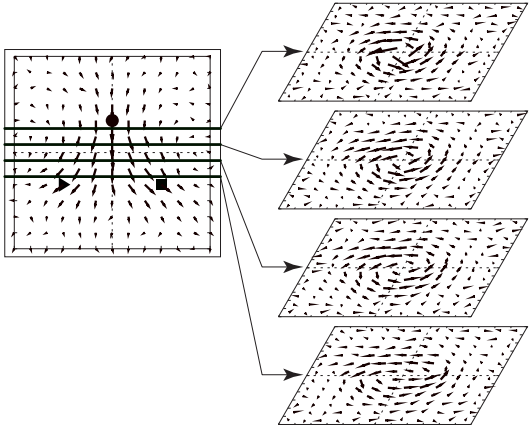}
\end{center}
\vspace{-9mm}
\caption{{\small \textit{Monopole current (right), obtained from the monopole component 
of the Abelian gauge field at $T/T_{c}=0.87$
}}}
\vspace{-7mm}
\label{fig:profile_B.eps}
\end{figure}
The same conclusions can be drawn from the Fig.\ref{fig:profile_S.eps} where 
the distribution of the monopole and photon parts of the action density is 
depicted.
Fig.\ref{fig:profile_B.eps} shows the monopole currents distribution.
One can see circulating monopole currents around the color electric field in each slice.
In the plane where the color electric field is divided into two parts, the circulating 
monopole current is not a perfect circle anymore.
This indicates a possibility of forming two circulating currents if the distance between 
quarks would be made larger.

\begin{figure}[thpb]
\begin{center}
\includegraphics[width=60mm]{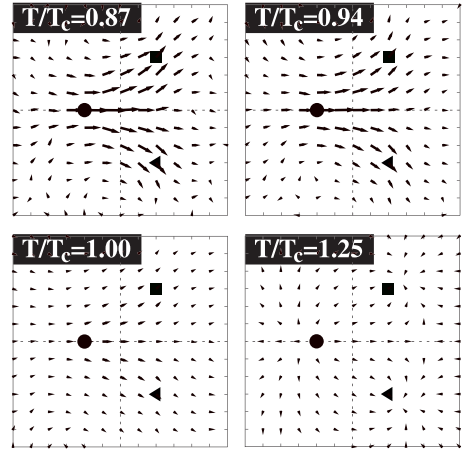}
\end{center}
\vspace{-9mm}
\caption{{\small \textit{Evolution of the color electric field (monopole component) 
with temperature.}}}
\vspace{-5mm}
\label{fig:profile_PT.eps}
\end{figure}

Next, we show how the profile of the flux tube changes when the temperature 
increases and crosses over to the high temperature phase.
From Fig.\ref{fig:profile_PT.eps} one can see that the squeezed color electric field (the monopole component) disappear at $T>T_c$.
In contrast, the photon component of the color electric field, depicted in
Fig.\ref{fig:profile_E.eps}(right), does not show any essential
changes  when the  temperature increases.
\section*{ACKNOWLEDGEMENTS}
This work is supported by the SR8000 Supercomputer Project
of High Energy Accelerator Research Organization (KEK).
A part of numerical measurements has been done using NEC SX-5
at RCNP of Osaka University.
T.S. is partially supported by JSPS Grant-in-Aid for Scientific Research on Priority Areas No.13135210 and (B) No.15340073.
The Moscow group is partially supported by RFBR  grants 02-02-17308, 01-02-17456, 00-15-96-786, grants INTAS--00-00111 DFG-RFBR 436 RUS 113/739/0, and CRDF awards RPI-2364-MO-02 and MO-011-0.
\vspace{-1mm}

\end{document}